\documentstyle[prb,aps,epsfig,preprint]{revtex}
\tightenlines
  
\newcommand{\be}{\begin{equation}}
\newcommand{\ee}{\end{equation}}

\newcommand{\ba}{\begin{eqnarray}}
\newcommand{\ea}{\end{eqnarray}}

\def\sg{{\sigma}}
\def\ros{{\rho(\sg)}}

\def\tro{{(T,[\rho])}}
\def\br{{\mathbf{r}}}

\begin{document}
\bibliographystyle{unsrt}
\title{Three-phase fractionation of polydisperse fluids}

\author{L. Bellier-Castella$^{\dag}$, M. Baus 
$^{\dag\dag}$ and H. Xu $^{\dag}$}
\date{April 24, 2001}
\maketitle

\noindent
$^{\dag}$ D\'epartement de Physique des Mat\'eriaux (UMR 5586 du CNRS),\\
Universit\'e Claude Bernard-Lyon1, 69622 Villeurbanne Cedex, France\\
$^{\dag\dag}$ Physique des Polym\`eres, Universit\'e Libre de Bruxelles,\\
Campus Plaine, CP 223, B-1050 Brussels, Belgium\\

\vspace{2truecm}
\pagebreak

\begin{abstract}
  It is shown that the  van der Waals free-energy of polydisperse
fluids, as introduced previously ( L. Bellier-Castella, H. Xu and M. Baus,
{\sl J. Chem. Phys.} {\bf 113}, 8337 (2000) ), predicts that for certain
thermodynamic states (e.g. low temperatures and large polydispersities) the ordinary two-phase coexistences become
metastable relative to a fractionation of the system into three
phases, reducing thereby the polydispersity of each of the 
coexisting phases. 
\end{abstract}

\pagebreak
\section{Introduction}
\label{sec1}
\vspace{1truecm}
A polydisperse system can be viewed as a ``continuous mixture",
i.e. as the limit of a discrete mixture whereby the number of
component species tends to infinity while the concentration
of each species tends to zero ~\cite{ref1}. Such a system
serves as a good model for many of the complex fluids
used in various industries ~\cite{ref2}. Whereas many of these
industrial fluids have badly characterized concentration distributions
one can find at present, e.g. in the physics laboratories, several
soft condensed matter systems ~\cite{ref3} (e.g. colloids, liquid
crystals, polymers, etc) with much better controled species distributions allowing
hence for a more detailed study of the influence of the polydispersity
on various physical properties (e.g. phase behavior, rheology, etc.).
To transpose the results and techniques elaborated for simple
fluids and their mixtures to polydisperse or continuous mixtures is
however by no means a simple task but has nevertheless recently become
an active field of research \cite{ref4}$^-$\cite{ref8}.\par
In the present study, a sequel to ref.8, we will be concerned with
the equilibrium phase behavior of polydisperse fluids. To simplify the
problem, we restrict ourselves to systems of spherical particles
(e.g. colloids) exhibiting a single polydispersity, say
a size-polydispersity. We assume moreover the initial (parent phase)
size-distribution to be monomodal, i.e. centered around a single reference
species, as suitable for the polydisperse generalization of
one-component systems. This initial size distribution will further
be assumed to be fixed, once and for all, by the production process of,
say, the colloidal particles. (Note that some systems, e.g. micellar
solutions, exhibit a ``variable" polydispersity whereby the size
distribution is allowed to adjust itself to some externally imposed
conditions \cite{ref9}.) When these colloidal particles interact by
excluded volume repulsions and, say, van der Waals (vdW) attractions
their thermodynamic properties can be studied on the basis of the
polydisperse generalization of the vdW free-energy \cite{ref10},
such as the one already used in our earlier studies \cite{ref7}$^-$\cite{ref8}.
Of course, such a description is far from being exact, but in general
the vdW-approximation captures, at least qualitatively, the essence
of the underlying phase behavior, as can be witnessed from several
previous investigations \cite{ref11}. As a final limitation,
we would like to stress that our study will be limited to the fluid
phases of the polydisperse system, leaving aside whether these phases
are stable or metastable with respect to the possible solid phases.
Indeed, the inclusion of the solid phases would require a more precise
specification of the interaction potential than is usual in a
vdW-description and, above all, would add still more complexity
to the already fairly complex problems raised by the study of
phase equilibria in polydisperse fluids. In this way, we will be
able to focus on the central difficulty resulting from the replacement of the
algebraic equations in finite dimensional spaces, characteristic of the
phase behavior of discrete mixtures, by the integral equations
in infinite dimensional spaces characteristic of the continuous mixtures.\par
In our previous study \cite{ref8}, the phase behavior of the present
system was already studied within the same vdW-approximation, but only
at the level of the two-phase coexistences. It was shown there that
the polydispersity can have a profound influence on the binodals of
this system as compared with its monodisperse counterpart. One aspect
which was not studied in ref.8 concerns the fact that, as a
consequence of the Gibbs phase rule, a polydisperse system can also
phase separate or ``fractionate" into more than two fluid phases, each
phase differing both in average density and in size distribution. The study
of such a polydispersity-induced equilibrium between three fluid phases
is the main object of the present investigation.
Meanwhile, we will also consider another question left unanswered
in ref.8. This concerns the specific form given to the parent-phase
 size-distribution. In ref.8 we used some well-known
expressions (Schulz-Zimm and log-normal) for this input
size-distribution. These theoretical distributions describe particles
with sizes ranging from zero to infinity. Since both
the very small and very large particles are absent from the experimentally
studied systems, one may wonder whether their presence in these
theoretical expressions could not lead to artifacts in the phase diagram.
The influence of these tails (for both small and large sizes) of the
theoretical distributions on the coexistence properties of ref.8
will hence be investigated here before proceeding to the three-phase
equilibria.\par
In section~\ref{sec2} we recall the vdW-description of polydisperse fluids as
used in ref.8. Section~\ref{sec3} is devoted to the influence of the
tails of the parent-phase distribution on the results of ref.8.
In section~\ref{sec4} we show that the present vdW-description allows for
the phase equilibrium between three fluid phases and we study the
relative stability of the two-phase and three-phase equilibria in section~\ref{sec5}.
Our conclusions are given in the final section~\ref{sec6}.\par
\section{The polydisperse vdW free-energy and its critical points}
\label{sec2}
The thermodynamic properties of a fluid of spherical particles
with a size-distribution can be determined from the following
generalization of the vdW free-energy (see ref.8 for details):
\ba
f(T,[\rho])=k_B\,T\int d\sg\ros\{\ln(
\frac{\Lambda^3(\sg)\ros}{E[\rho]})-1\}\nonumber\\
+\frac{1}{2}\int d\sg\int d\sg'\,V(\sg,\sg')\ros\rho(\sg')
\label{eq1}
\ea
where, $f(T,[\rho])$, is the free-energy per unit volume at the temperature
$T$ ($k_B$ being Boltzmann's constant) for a fluid for which the number
density of species $\sg$ is $\ros$, while $\Lambda(\sg)$ denotes the
thermal de Broglie wavelength of species $\sg$. Any functional dependence
on the density distribution $\ros$ is, as usual, indicated by $[\rho]$.
In (\ref{eq1}), the excluded volume repulsions are represented by:
\be 
E[\rho]=1-\int d\sg\,v(\sg)\ros
\label{eq2}
\ee
where, $v(\sg)=\frac{4\pi}{3}\left(R(\sg)\right)^3$, is the volume
of a spherical particle of radius $R(\sg)$, whereas the attractions are
represented by:
\be
V(\sg,\sg')=\int d\br\,V_A(r;\sg,\sg').
\label{eq3}
\ee
with $V_A(r;\sg,\sg')$ being the potential of attraction between
two particles of species $\sg$ and $\sg'$, a distance $r=|\br|$ apart.
In the above $\sg$ represents both a species label and the dimensionless
polydispersity variable, $R(\sg)/R(1)$, $R(1)$ being the radius of the
reference species, $\sg=1$.\par
The pressure, $p\tro$, corresponding to (\ref{eq1}) is:
\be
p\tro=\frac{k_B\,T}{E[\rho]}\int d\sg\ros
+\frac{1}{2}\int d\sg\int d\sg'\,V(\sg,\sg')\ros\rho(\sg').
\label{eq4}
\ee
whereas the chemical potential of species $\sg$, $\mu(\sg,T,[\rho])$,
reads:
\ba
\mu(\sg,T,[\rho])=k_B\,T\ln\{
\frac{\Lambda^3(\sg)\ros}{E[\rho]}\}
+k_B\,T\frac{v(\sg)}{E[\rho]}\int d\sg'\rho(\sg')\nonumber\\
+\int d\sg'\,V(\sg,\sg')\rho(\sg').
\label{eq5}
\ea
In the above, all integrals over $\sg$ extend over the whole domain for
which $\ros$ is non-zero. As in ref.8, we write, $\ros=\rho\,h(\sg)$,
where $\rho=\int d\sg\ros$ is the average density and $h(\sg)$ the
(normalized) size-distribution. The dimensionless average density will be
written as, $\eta=\rho\,v(1)$, with $v(1)$ the volume of the reference
species $\sg=1$, while the dimensionless temperature will be,
$t=k_B T/\epsilon(1,1)$, with $\epsilon(1,1)=-V(1,1)/8v(1)$ and
$V(1,1)$ being the integrated amplitude of the attraction between two
reference particles (cf.(\ref{eq3}) and ref.8).\par
In ref.8 we have considered several types of polydisperse interactions
characterized by two parameters \{$l,n$\}, viz.
$v(\sg)=v(1)\sg^{3n}$ and $V(\sg,\sg')=[(\sg^n+\sg'^n)/2]^3(\sg\sg')^l V(1,1)$.
As shown there the model with \{$l=1,n=0$\} has a phase behavior which is
similar to that with the full polydispersity \{$l=1,n=1$\} but is
simpler to study because its excess free-energy involves fewer moments,
$m_k=\int d\sg\sg^k h(\sg)$, of the size-distribution $h(\sg)$. Henceforth,
we will therefore consider only this \{$l=1,n=0$\}-model, so as to
simplify the calculations. In ref.8 we also considered several
types of parent-phase size-distributions $h_0(\sg)$ but found
no qualitative differences between them. In order to avoid too much
duplication of results, we will henceforth consider only 
the log-normal parent-phase size-distribution, 
$h_0(\sg)=c\;\exp[-a\,ln^2(\sg/b)]$, where the three parameters
\{$a,b,c$\} are determined by imposing the first three moments
of $h_0(\sg)$, $m_k^{(0)}=\int_0^\infty d\sg \sg^k h_0(\sg)$,
to be such that: 1) $h_0(\sg)$ be normalized ($m_0^{(0)}=1$),
2) the average value of $\sg$ be one, i.e. equal to the reference
species ($m_1^{(0)}=1$), 3) the variance of $h_0(\sigma)$ 
be
$\frac{1}{\alpha}=I-1>0$, with $I$ being the polydispersity
index ($m_2^{(0)}=I$). In terms of $I$ this yields: $a=1/2lnI$,
$b=I^{-3/2}$, $c=I/\sqrt{2\pi lnI}$ and $m_k^{(0)}=I^{k(k-1)/2}$.\par
When $I=1$ (or $\alpha=\infty$) the system is monodisperse
($h_0(\sg)=\delta(\sg-1)$) and its phase diagram consists of the
usual vdW-binodal ending in the vdW critical point ($\eta_c=1/3$,
$t_c=32/27$). For a modest polydispersity, say $I=1.02$ or
$\alpha=50$, the phase diagram is similar to that of Fig.4 of
ref.8. For each parent-phase density $\eta_0$, there now is
a different binodal. Each binodal is truncated upwards at a maximum
temperature, $t_m$, with the corresponding densities,
$\eta_1(t_m)$ and $\eta_2(t_m)$, lying respectively on the
cloud-point and shadow curves. For a critical value of $\eta_0$,
$\eta_0=\eta_c$, the corresponding binodal passes through the
intersection of the latter two curves. This occurs for
$t=t_c$ and, since, $\eta_1(t_c)=\eta_2(t_c)$, the point
($\eta_c,t_c$) is a critical point where the two coexisting
phases ($1$ and $2$) become identical. However, in contradistinction
with the monodisperse case ($I=1$) the critical temperature
is not the borderline between the one-phase and two-phase
regions. Indeed, in the polydisperse case, the two-phase region
extends to $t'_m>t_c$, where $t'_m$ corresponds to the maximum
of the cloud-point and shadow curves. At the same time the critical
point loses one of its attributes, i.e. for $I\ne 1$, there also
appears a second high-density critical point. The latter is
polydispersity-induced \cite{ref8} but, as will be shown below,
not necessarily thermodynamically stable. When the polydispersity
index $I$ is increased still further the ordinary low-density
(vdW) critical point moves to a higher density and a lower
temperature, while the polydispersity-induced high-density
critical point moves to a lower density and higher temperature,
until for a limiting value of $I$, say $I=I^*$, the two
critical points merge as illustrated in Fig.1. For $I>I^*$
there are no critical points and the phase behavior loses all
contact with its monodisperse ($I=1$) counterpart.\par
\section{Truncated parent-phase size-distribution}
\label{sec3}
For a polydisperse system the input data for the study of its
phase behavior involve, besides the temperature ($t$) and
the average density ($\eta_0$), the parent-phase size-distribution
$h_0(\sg)$ (as in ref.8 the subscript zero refers to the
parent-phase). In theoretical work the values of $\sg$ are
usually distributed over the full interval of all possible
sizes, $0\leq\sg\leq\infty$,
whereas in the samples used in the laboratory these sizes usually
span a continuous but finite interval, $\sg^{\star}\leq\sg\leq\sg^{\star \star}$,
with $0<\sg^{\star}<1$ and $1<\sg^{\star \star}<\infty$ if $\sg=1$ represents
the reference size (usually taken to be the average size). As
stated already in ref.8, the small-size ($0<\sg<\sg^{\star}$) and
the large size ($\sg^{\star \star}<\sg<\infty$) tails of $h_0(\sg)$ will
be harmless for strongly peaked distributions, i.e. for small
polydispersities, but, as stated in the Introduction, they could
lead to artifacts for large polydispersities characterized by
broad size-distributions. Before studying the phase behavior of
our system in the region of the (large) limiting polydispersity,
$I\approx I^*$, where any contact with the monodisperse system is
lost, we will first assess the influence of these tails on the results
of ref.8. To this end we again consider the log-normal distribution
of section~\ref{sec2} but normalize it now over the finite interval 
($\sg^{\star},\sg^{\star \star}$) and rescale $\sg$
in such a manner that the average value of $\sg$ computed over
this finite interval be again equal to one, i.e. the reference
particle remains unchanged. In Fig.2 we show the effect of
this truncation of the $\sg$-domain
on the form of $h_0(\sg)$ for some representative values of
\{$\sg^{\star},\sg^{\star \star}$\}. In Table 1 we show its influence on the thermodynamic
data. The overall effect of this truncation is hence to render
the system less polydisperse. It also shifts the coexistence
densities to lower values. Since some of our coexisting phases
have densities for which the fluid phases could become metastable
with respect to solid phases, a truncation of $h_0(\sg)$ could render these
fluid phases their thermodynamic stability. In any case,
a quantitative comparison with experimental results will require
phase diagrams calculated for truncated
size-distributions and hence require information about $\sg^{\star}$ and $\sg^{\star \star}$.
Since, however, no qualitative differences (or artifacts) are found,
we will continue henceforth with the ``untruncated" $h_0(\sg)$
log-normal distribution, as defined at the end of section~\ref{sec2}
.\par
\par

\section{Three-phase equilibria: local stability}
\label{sec4}
As seen in section~\ref{sec2}, for a sufficiently large polydispersity index
($I>I^{\star}$) the phase behavior of our system loses any contact
with its monodisperse counterpart. This is hence a good region to
look for novel features such as three-phase equilibria. In the
present section we will investigate how an initial (untruncated)
log-normal parent-phase size-distribution, $h_0(\sg)$, fractionates
into (two or three) daughter phases when its polydispersity index
$I$ lies on either side of the limiting value $I^*$ ($I^*=1.072$ or
$\alpha^*=13.72$) where the critical points disappear. We will focus
on two cases: $\alpha=15$ or $I=1.066$ and $\alpha=13.5$ or
$I=1.074$.\par
\subsection{Two-phase equilibria}
\label{sec4a}
As discussed in detail in ref.8, these can be obtained by
solving the two-phase coexistence conditions:
\be
p(T,[\rho_1])=p(T,[\rho_2]) ;
\;\;\mu(\sg,T,[\rho_1])=\mu(\sg,T,[\rho_2]) 
\label{eq6}
\ee
where \{$\rho_1(\sg)=\rho_1h_1(\sg)$, $\rho_2(\sg)=\rho_2h_2(\sg)$\}
are the density distributions of the two daughter-phases resulting from
the fractionation of the parent-phase, $\rho_0(\sg)=\rho_0h_0(\sg)$.
The integral equation resulting from (\ref{eq6}) can be solved as explained
in ref.8. For the large $I$-values considered here we find
two sets of binodals, each one being linked to a different critical point
(see section~\ref{sec2}). Since at these critical points the system is only
marginally stable, i.e. we have $\delta^2 f(T,[\rho])=0$, we first
inquire for the local stability ($\delta^2 f>0$) of these
solutions with respect to an infinitesimal change $\delta\rho(\sigma)$ of $\ros=\rho_n(\sigma)$ ($n=1,2$):
\be
\delta^2 f=\int d\sg_1\int d\sg_2\; 
\frac{\delta^2 f(T,[\rho])}{\delta\rho(\sg_1)\delta\rho(\sg_2)
}\;\delta\rho(\sg_1)\delta\rho(\sg_2)\;\;>0
\label{eq7}
\ee
(see ref.8 for the explicit form of $\delta^2f$). It turns out
that, for the $\eta_0$-values investigated, one of the two sets
is locally unstable, i.e. it does not satisfy (\ref{eq7}) (see Fig.3).\par
\subsection{Three-phase equilibria}
\label{sec4b}
Similarly, the phase equilibria between three fluid-phases are governed by the solutions of :\par
\ba
p(T,[\rho_1])=p(T,[\rho_2])=p(T,[\rho_3]) \nonumber\\
\mu(\sg,T,[\rho_1])=\mu(\sg,T,[\rho_2])=\mu(\sg,T,[\rho_3])
\label{eq8}
\ea
where \{$\rho_1(\sg)=\rho_1h_1(\sg)$, $\rho_2(\sg)=\rho_2h_2(\sg)$,
$\rho_3(\sg)=\rho_3h_3(\sg)$\} are the density distributions of the
three daughter-phases. To solve (\ref{eq8}) we can generalize
the method used in ref.8. The $h_n(\sg)$ ($n=1,2,3$) are
constrained by the particle number conservation of species $\sg$,
$x_1h_1(\sg)+x_2h_2(\sg)+x_3h_3(\sg)=h_0(\sg)$, whereas the total
particle number conservation implies $x_1+x_2+x_3=1$, $x_n$ being
the fraction of the parent-phase particles which went into phase
$n$ (see ref.8 for details). The $\rho_n$ are constrained by
the conservation of the total volume, $x_1v_1+x_2v_2+x_3v_3=v_0$,
with $v_n=1/\rho_n$ ($n=0,1,2,3$). This leaves us with two independent
distributions, say $h_1(\sg)$ and $h_2(\sg)$, two independent
number fractions, say $x_1$ and $x_2$, and two independent
number densities, say $\rho_1$ and $\rho_2$. Given ($x_1,x_2$) and
($\rho_1,\rho_2$), the two distributions, $h_1(\sg)$ and $h_2(\sg)$,
can be found by solving the system of two integral equations resulting
from $\mu(\sg,T,[\rho_1])=\mu(\sg,T,[\rho_2])$ and
$\mu(\sg,T,[\rho_2])=\mu(\sg,T,[\rho_3])$. As in ref.8, these
two equations can be rewritten as: $h_1(\sg)=h_0(\sg)H_1(\sg)$
and $h_2(\sg)=h_0(\sg)H_2(\sg)$, so that the normalization of the $h_n(\sigma)$ can be expressed as, $1=\int d\sigma h_0(\sigma) H_1(\sigma)$ and $1=\int d\sigma h_0(\sigma) H_2(\sigma)$, whereas solving the latter
two equations together with, $p(T,[\rho_1])=p(T,[\rho_2])$
and $p(T,[\rho_2])=p(T,[\rho_3])$ will determine ($x_1,x_2$) and
($\rho_1,\rho_2$). The full solution of a three-phase equilibrium
problem for a polydisperse fluid is thus a rather complex task
(hereby justifying some of the simplifying assumptions introduced
above). In the case where the excess free-energy depends only on a finite
number of moments of the distribution $h(\sg)$, as is the case
for the vdW free-energy, the above procedure can be simplified
as the integral equations can then be transformed into a finite
set of non-linear relations between these moments, as explained in
ref.8. We found that the amount of labor involved is similar to
that of the projection method of ref.6 but contrary to the
latter, the present method involves no approximation
to the basic equations (\ref{eq6}) or (\ref{eq8}). In order to
speed up the convergence of the solution method one may use the
Powell-algorithm \cite{ref12} instead of the more traditional Newton-Raphson
method \cite{ref13} used in ref.8. An example of a three-phase
coexistence found in this way is shown in Fig.4. We have verified that
the case shown is locally stable, i.e. the $\rho_n(\sigma)$ ($n=1,2,3$)
satisfy (\ref{eq7}).\par
\section{Relative stability of the two-phase and three-phase
equilibria}
\label{sec5}
As seen in the previous section, in strongly polydisperse
systems it is possible to obtain, for the same input data,
 locally stable two-phase as well
as three-phase solutions. This then raises
the question of how to separate the stable from the metastable
transitions, i.e. of the global stability of these solutions. All
the globally stable states belong to the convex envelope of the
free-energy surface. This envelope is such that the tangent plane
through any of its points never cuts the free energy surface. Of
course, in the polydisperse case these ``surfaces" are defined
in the infinite dimensional functional space supporting the density
distributions, $\ros$. As a consequence, if $\rho_e(\sg)$ belongs to
the convex envelope of $f(T,[\rho])$ it must satisfy:
\be
f(T,[\rho])\geq f(T,[\rho_e])+\int d\sg\frac{\delta f(T,[\rho_e])}
{\delta \rho_e(\sg)}\;(\ros-\rho_e(\sg))
\label{eq9} 
\ee
for any $\ros$. It is of course not possible to verify this infinite
number of conditions but, in practice, it will suffice to verify
(\ref{eq9}) for those states $\rho_e(\sg)$ which are candidates
for a stable equilibrium, i.e. for the solutions of eqs. (\ref{eq6}) or
(\ref{eq8}), and for those states $\ros$ for which the distance
between the free-energy surface and the tangent plane through
$\rho_e(\sg)$ is extremal, i.e. for the $\rho(\sigma)$ solution of:
\be
\frac{\delta f(T,[\rho])}{\delta \rho(\sg)}
=\frac{\delta f(T,[\rho_e])}{\delta \rho_e(\sg)}
\label{eq10}
\ee 
which usually amount to a finite
number \cite{ref6} for each $\rho_e(\sg)$. In doing so, it is
of course still possible that there exist (in the same
parameter region) higher-order equilibria involving four, five, etc.
phases which could still invalidate (\ref{eq9}), but in view of the
tremendous complexity of the global stability problem, we will
limit ourselves here to the relative stability of the two-phase
and three-phase equilibria found thus far. As a result of this
search we find that the locally stable two-phase equilibrium
solution of (\ref{eq6}) is globally stable between an upper
($t_+$) and a lower ($t_-$) temperature. Above the upper temperature
($t>t_+$) the parent-phase is stable, while for $t<t_+$ the parent
-phase first fractionates into two phases while for $t<t_-$ the
system further fractionates into three phases. Two examples of
such phase diagrams are shown in Fig.5 and Fig.6.\par
\section{Conclusions}
\label{sec6}
The investigation of the phase behavior of the polydisperse
generalization of the vdW free-energy which was started
in ref.8 for relatively modest polydispersities ($I\approx1.02$)
has been extended here to larger polydispersities ($I\approx 1.07$).
It is seen that when the polydispersity is increased the system's phase
behavior gradually loses contact with its monodisperse
counterpart. Indeed, for small polydispersities, the system
investigated has two critical points, one of which is the polydisperse
generalization of the vdW critical point, the other being
polydispersity-induced. When the polydispersity is increased
these two critical points merge for a limiting polydispersity
($I^*\approx1.072$). Above this threshold value ($I^*$) there
are no critical points. Similarly, when the polydispersity
is increased, the two-phase region splits
into two two-phase regions for the higher temperatures whereas for the lower temperatures a polydispersity-induced three-phase region appears.
In order to determine the boundaries between these different regions
of the phase diagram it is essential to investigate both the local and
global stability of a large number of possible phase transitions.
While some of these transitions are not even locally stable others
are locally and globally stable. The situation quickly becomes very
complex both physically and mathematically.\par
The phase behavior investigated here should be of relevance to
the study of colloidal dispersions of spherical particles for which
the size-distribution is monomodal. A detailed comparison will of course
require a further extension of the present investigation
towards the solid phases, in particular their thermodynamic
stability \cite{ref14} for the large polydispersities
considered here.\par
\vspace{0.3truecm}

\noindent
{\bf Acknowledgements}
\newline
M.B. acknowledges financial support from the F.N.R.S.

\pagebreak
\pagebreak
\noindent{\bf Figure Captions}\par
\vspace{1truecm}
\noindent
{\bf FIG. 1.} The low-density (circles) and high density (squares) critical points versus the polydispersity ($I=1+\frac{1}{\alpha}$). The limiting value of $\alpha$ below which the two critical points disappear is $\alpha^{\star}=13.72$ ($I^{\star}=1.073$). Fig. 1a (resp. Fig. 1b)  displays $\eta_c$ (resp. $t_c$) versus $\alpha$.  \par
\vspace{0.3truecm}
\noindent
{\bf FIG. 2.} The full log-normal size distribution (dashed-dotted line) compared to a log-normal distribution truncated for $\sigma<\sigma^{\star}$ and $\sigma^{\star \star}<\sigma$ (full line). Both distributions are normalized ($m_0=1=m_0^t$) and have the same average $\sigma$-value ($m_1=1=m_1^t$). The case shown here corresponds to $\alpha=15$ ($I=1.067$) and $\sigma^{\star}=0.72$, $\sigma^{\star \star}=1.33$.\par
\vspace{0.3truecm}
\noindent
{\bf FIG. 3.} Two sets of binodals for $\alpha=15$ and $\eta_0=0.68$. The full-line binodals correspond to locally stable solutions of eq.(6) while the dotted-line binodals are locally unstable. Also represented (diamond) is the second (unstable) critical point.\par
\vspace{0.3truecm}
\noindent
{\bf FIG. 4a} A three-phase equilibrium for $\alpha=15$ and $\eta_0=0.506$ corresponding to a locally stable solution of eq.(8). The coexisting densities at $t=1.07$ are $\eta_1=0.284$, $\eta_2=0.652$ and $\eta_3=0.930$.\par
\vspace{0.3truecm}
\noindent
{\bf FIG. 4b} Coexisting distributions $h_i(\sigma)$ ($i=1$: full-line, $i=2$: dotted-line and $i=3$: dashed-dotted line) at $t=1.07$ and for the $\alpha$ and $\eta_0$ values of Fig. 4a.\par
\vspace{0.3truecm}
\noindent
{\bf FIG. 5} A phase diagram for $\alpha=15$ and $\eta_0=0.506$. The one-phase region corresponds to $t>t_{+} \simeq 1.75$, the two-phase region to $t_{+} <t <t_{-} \simeq 1.24$, while the three-phase region corresponds to $t<t_{-}$.\par
\vspace{0.3truecm}
\noindent
{\bf FIG. 6} The same as Fig.5 but for $\alpha=13.5$ and $\eta_0=0.506$. Here $t_{+}\simeq 1.75$ and \\
$t_{-}\simeq 1.26$. Note that while in Fig.5 the new (third) phase is a low-density phase here it is an intermediate-density phase.

\vspace{0.5cm}
\begin{center}
\begin {table}
\( \begin{array} {|c|c|c|c|} \hline
\sigma^{\star}&\sigma^{\star \star}&\eta_1&\eta_2\\ \hline
  0  &  \infty  &  0.24  &  0.69  \\ \hline
0.30&2.75&0.24&0.69\\ \hline
0.48&1.76&0.23&0.68\\ \hline
0.56&1.54&0.20&0.66\\ \hline
0.70&1.32&0.15&0.63\\ \hline
\end{array} \)
\vspace{0.5cm}
\caption{Typical shifts in the coexisting densities of the low-density ($\eta_1$) and high density ($\eta_2$) phase for $t=1$, $\alpha=15$ after truncation of the log-normal parent-phase size-distribution for $\sigma<\sigma^{\star}$ and $\sigma^{\star \star}< \sigma$. The corresponding monodisperse results are $\eta_1=0.10$, $\eta_2=0.61$.}
\end{table}
\end{center}
\newpage
\thispagestyle{empty}
\hspace{5cm}
{\Large
\textbf{
Fig. 1a}
, Bellier-Castella et al., JCP}
\\[2cm]
\Large
\begin{center}
\unitlength1mm
\begin{picture}(140,140)
\put(0,0){\makebox(140,140)[b]
{\epsfysize140mm \leavevmode \epsffile{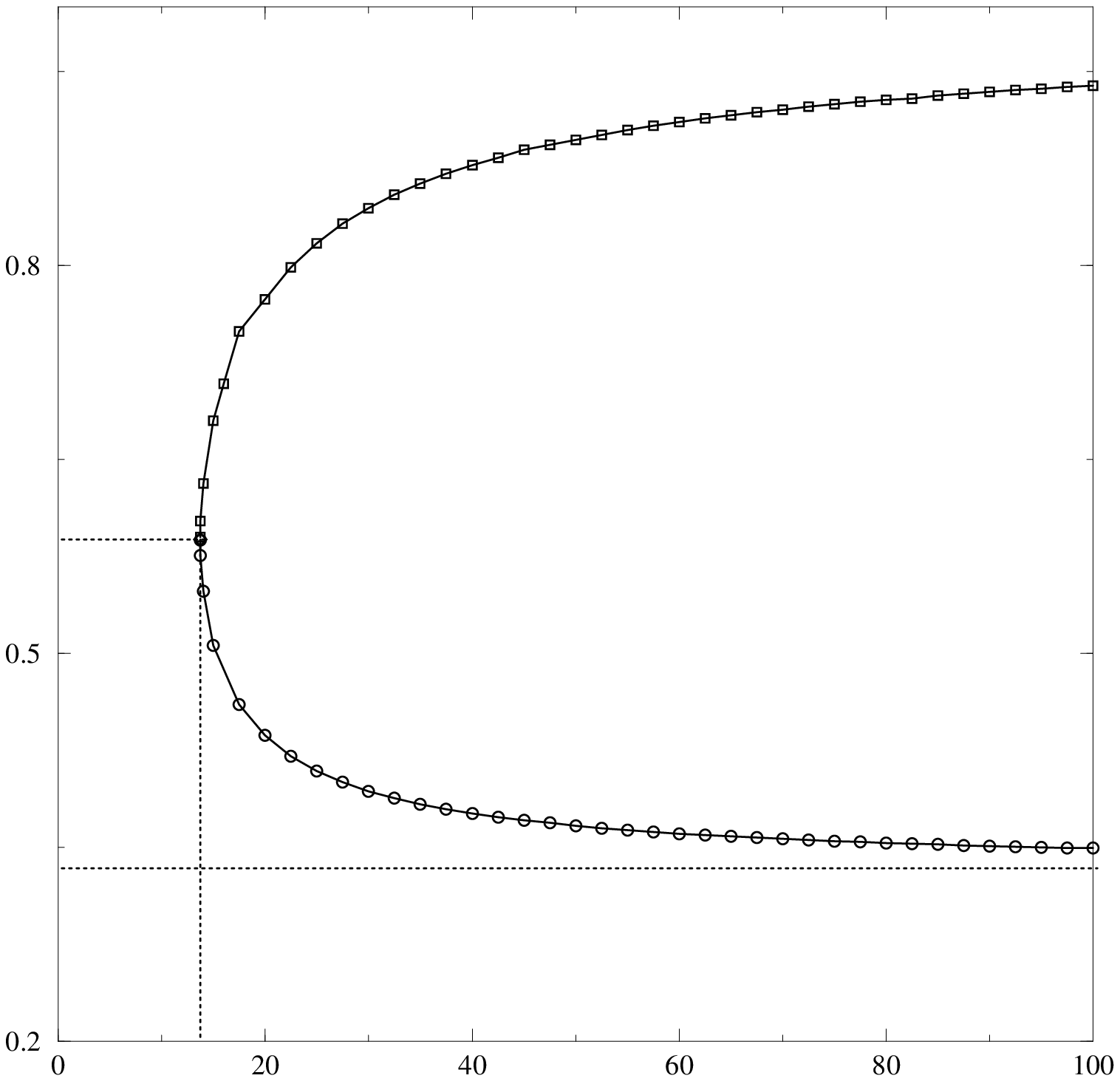} }}
\end{picture}
\end{center}

\vskip -1 cm
\hskip 8 cm
{\Large
$\alpha$}

\vskip -1.5 cm
\hskip 2.75 cm
{\large
$\alpha^{\star}$}

\vskip -10. cm
\hskip -0.5 cm
{\Large
$\eta_c$}

\vskip 0.15 cm
\hskip -0.5 cm
{\large
$0.59$}

\vskip 3. cm
\hskip -0.5 cm
{\large
$1/3$}

\newpage
\thispagestyle{empty}
\hspace{5cm}
{\Large
\textbf{
Fig. 1b}
, Bellier-Castella et al., JCP}
\\[2cm]
\Large
\begin{center}
\unitlength1mm
\begin{picture}(140,140)
\put(0,0){\makebox(140,140)[b]
{\epsfysize140mm \leavevmode \epsffile{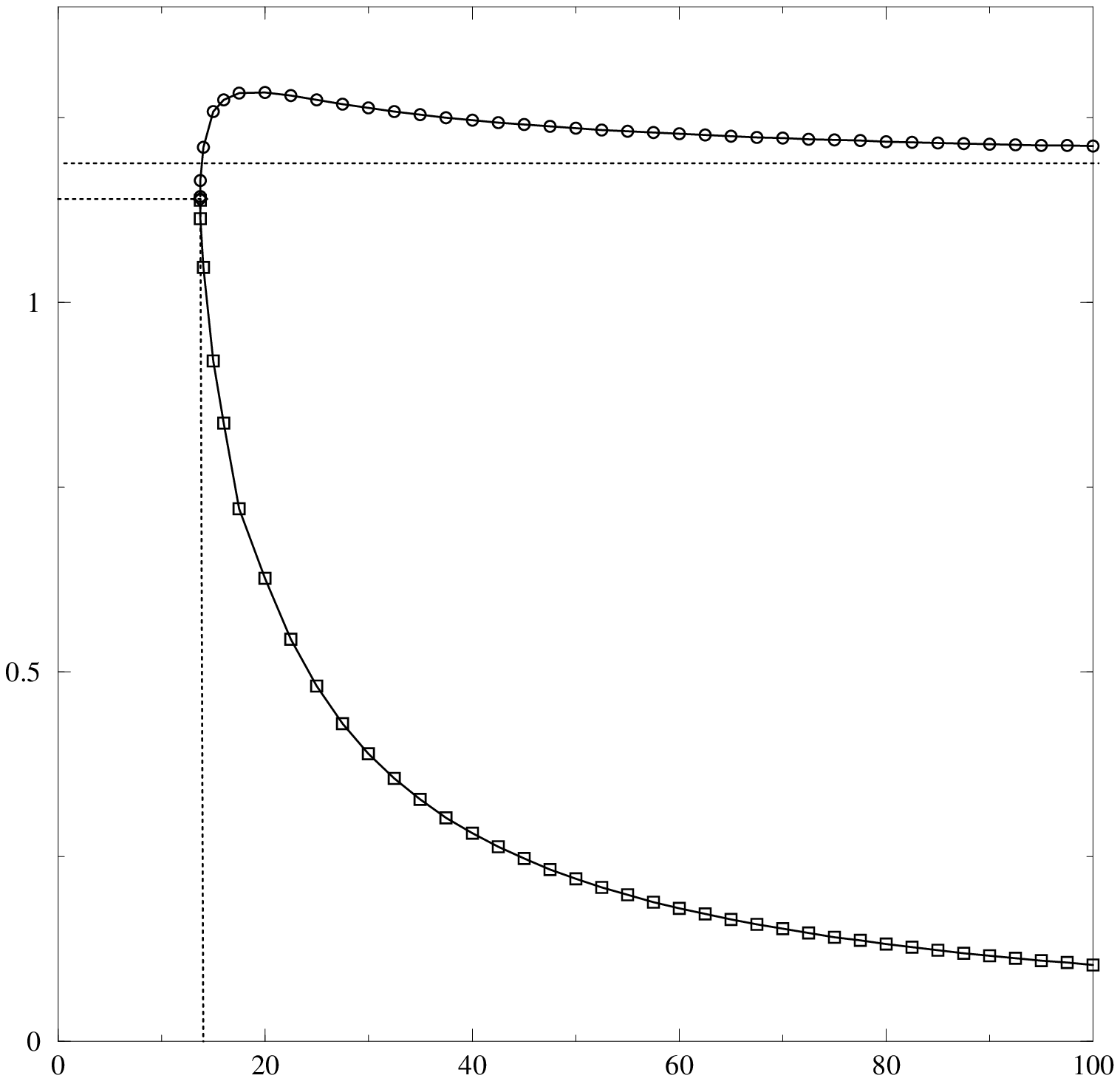} }}
\end{picture}
\end{center}

\vskip -1 cm
\hskip 8 cm
{\Large
$\alpha$}

\vskip -1.5 cm
\hskip 2.7 cm
{\large
$\alpha^{\star}$}

\vskip -10. cm
\hskip -0.5 cm
{\Large
$t_c$}

\vskip -4.3 cm
\hskip -0.5 cm
{\large
$1.14$}

\vskip -1.75 cm
\hskip -0.5 cm
{\large
$1.18$}

\newpage
\thispagestyle{empty}
\hspace{5cm}
{\Large
\textbf{
Fig. 2}
, Bellier-Castella et al., JCP}
\\[2cm]
\Large
\begin{center}
\unitlength1mm
\begin{picture}(140,140)
\put(0,0){\makebox(140,140)[b]
{\epsfysize140mm \leavevmode \epsffile{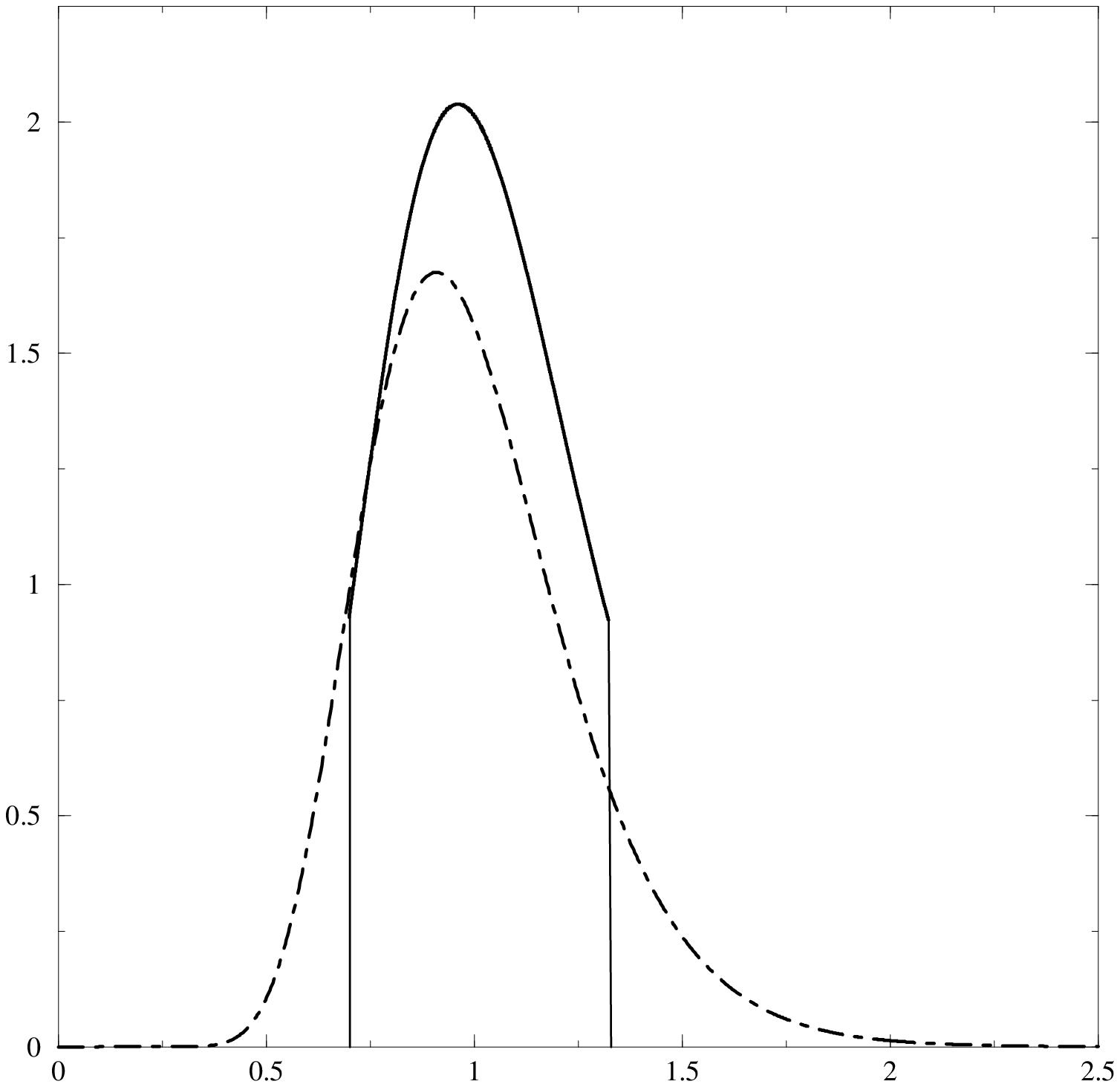} }}
\end{picture}
\end{center}

\vskip -1 cm
\hskip 11 cm
{\Large
$\sigma$}

\vskip -1.5 cm
\hskip 4.65 cm
{\large
$\sigma^{\star}$}

\vskip -1.25 cm
\hskip 8. cm
{\large
$\sigma^{\star \star}$}


\vskip -12. cm
\hskip -1. cm
{\Large
$h_0(\sigma)$}

\newpage
\thispagestyle{empty}
\hspace{5cm}
{\Large
\textbf{
Fig. 3}
, Bellier-Castella et al., JCP}
\\[2cm]
\Large
\begin{center}
\unitlength1mm
\begin{picture}(140,140)
\put(0,0){\makebox(140,140)[b]
{\epsfysize140mm \leavevmode \epsffile{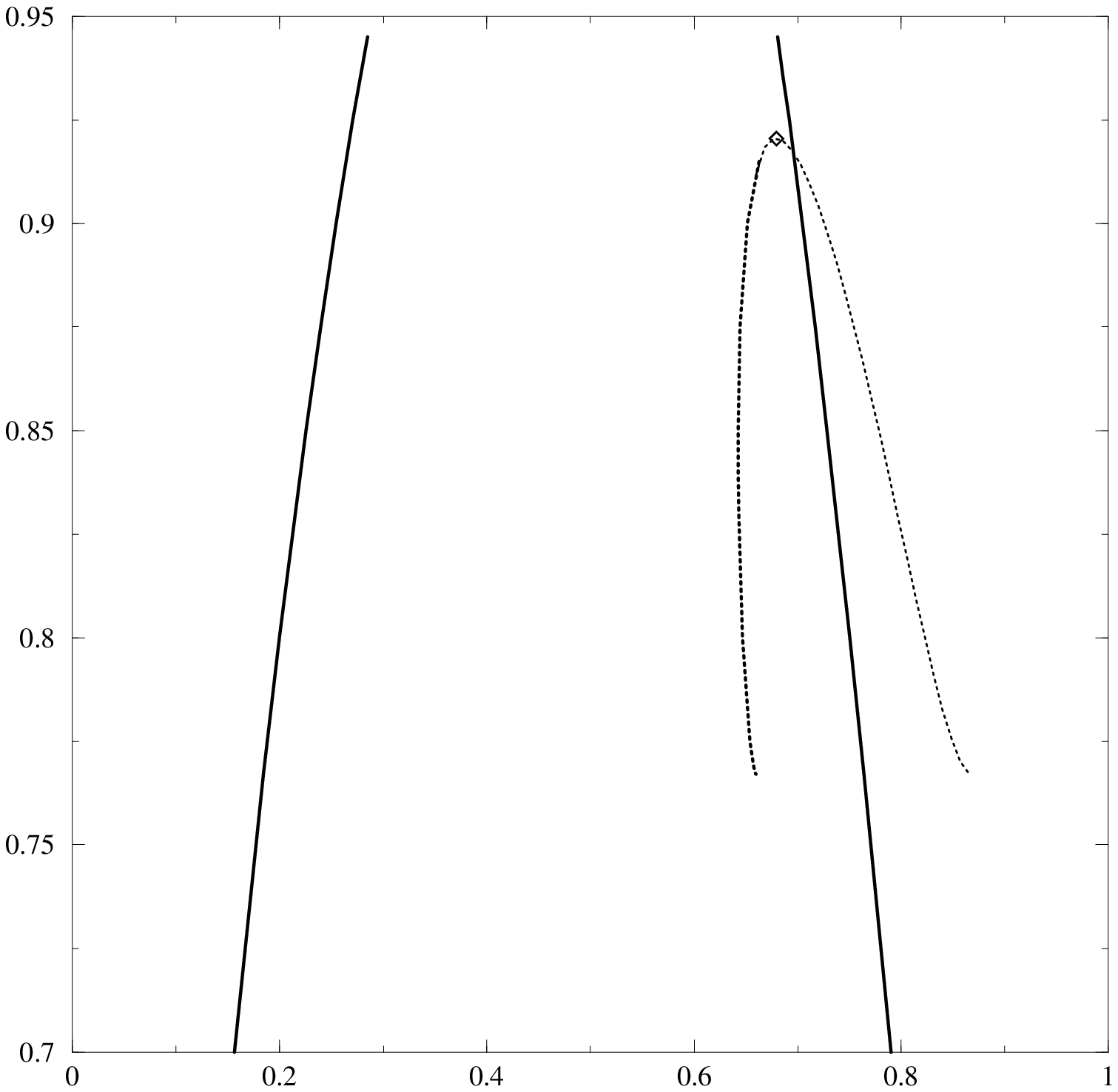} }}
\end{picture}
\end{center}

\vskip -1 cm
\hskip 12 cm
{\Large
$\eta$}

\vskip -12 cm
\hskip 0. cm
{\Large
$t$}

\newpage
\thispagestyle{empty}
\hspace{5cm}
{\Large
\textbf{
Fig. 4a}
, Bellier-Castella et al., JCP}
\\[2cm]
\Large
\begin{center}
\unitlength1mm
\begin{picture}(140,140)
\put(0,0){\makebox(140,140)[b]  
{\epsfysize140mm \leavevmode \epsffile{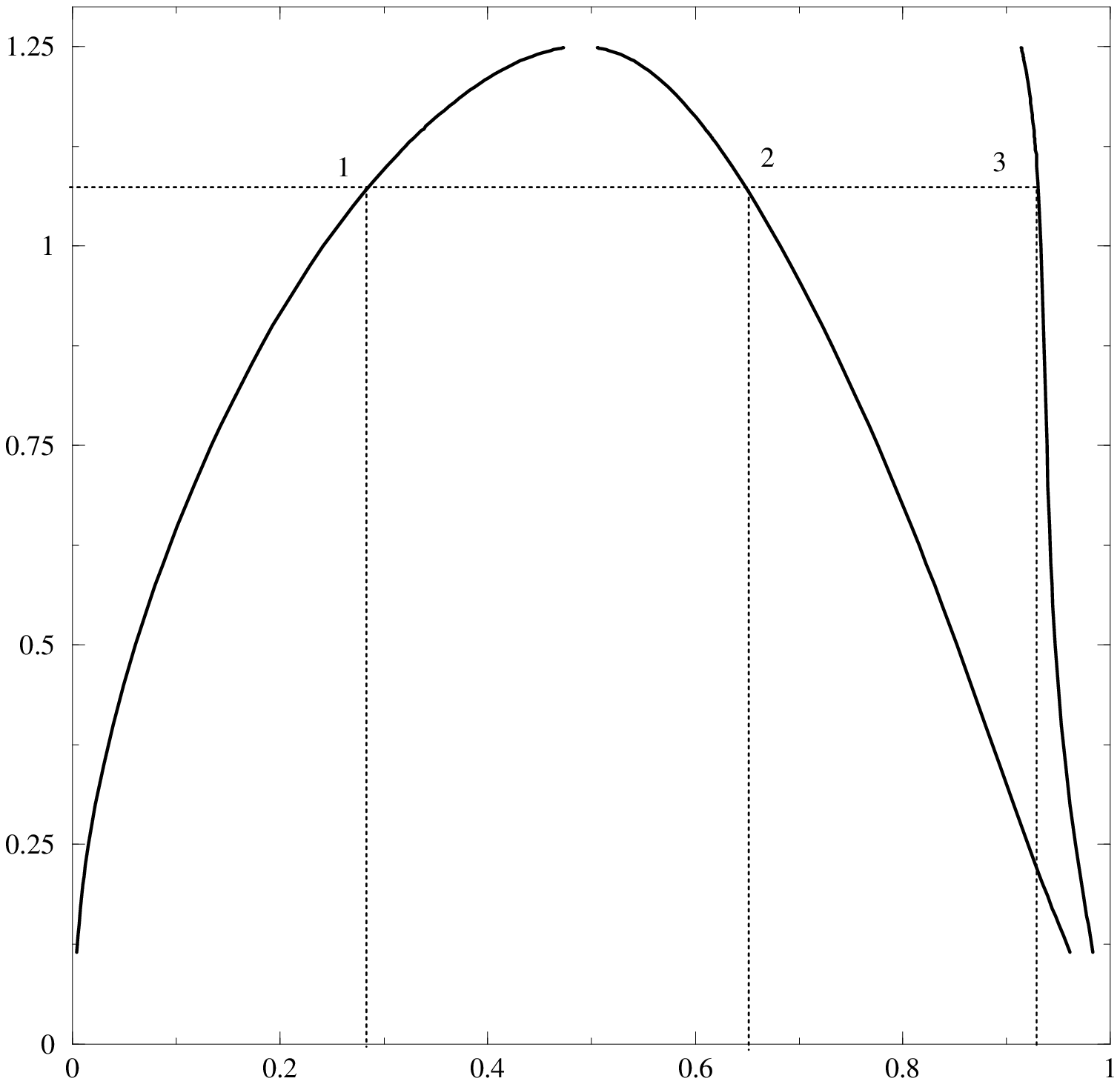} }}
\end{picture}
\end{center}

\vskip -1 cm
\hskip 12 cm
{\Large
$\eta$}

\vskip -12 cm
\hskip 0. cm
{\Large
$t$}

\vskip 9 cm
\hskip 4.75 cm
{\large
$\eta_1$}

\vskip -1.25 cm
\hskip 9.75 cm
{\large
$\eta_2$}

\vskip -1.25 cm
\hskip 13.5 cm
{\large
$\eta_3$}

\vskip -13. cm
\hskip 0. cm
{\large
$1.07$}
\newpage
\thispagestyle{empty}
\hspace{5cm}
{\Large
\textbf{
Fig. 4b}
, Bellier-Castella et al., JCP}
\\[2cm]
\Large
\begin{center}
\unitlength1mm
\begin{picture}(140,140)
\put(0,0){\makebox(140,140)[b]  
{\epsfysize140mm \leavevmode \epsffile{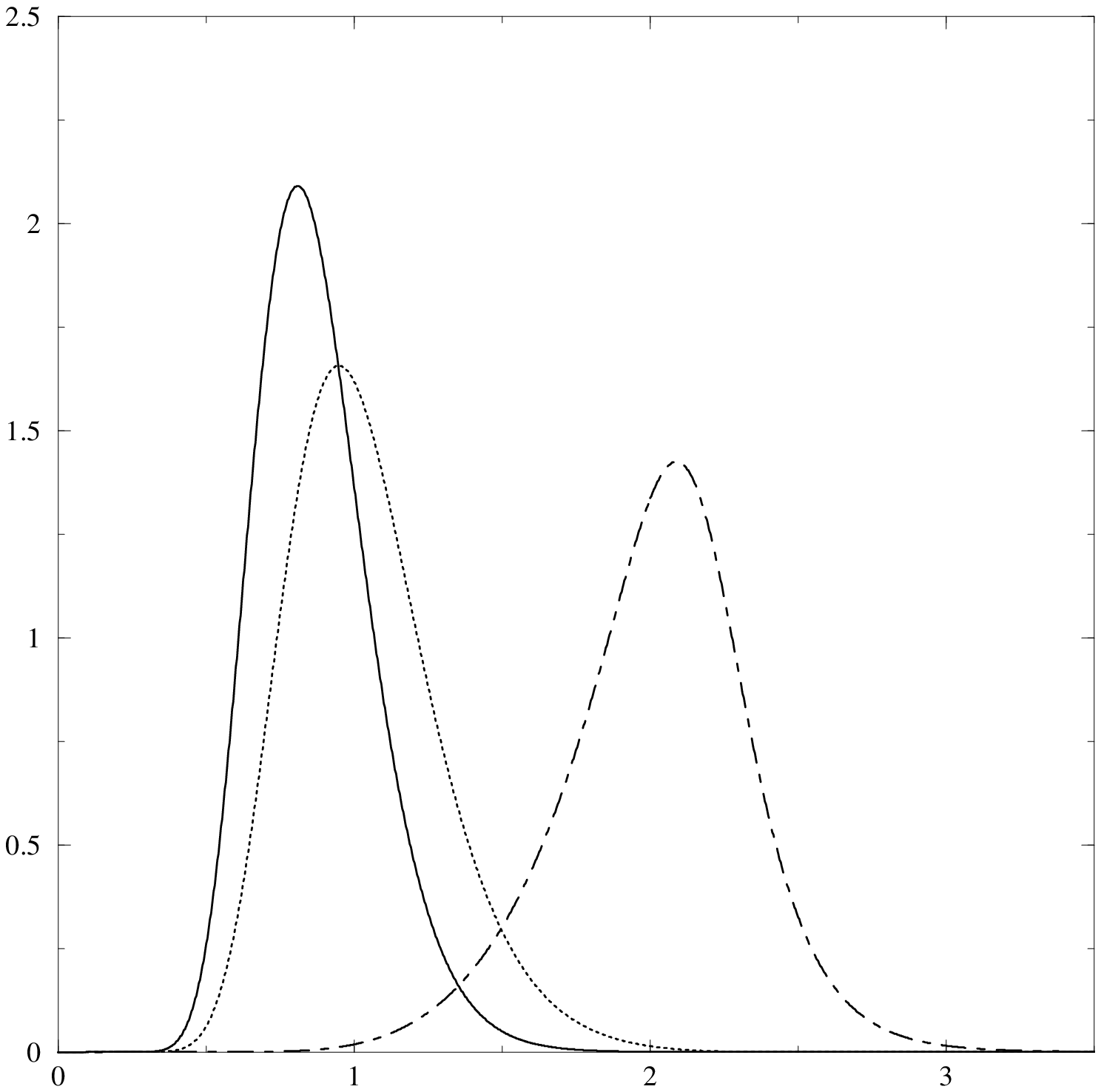} }}
\end{picture}
\end{center}

\vskip -1 cm
\hskip 12 cm
{\Large
$\sigma$}

\vskip -12 cm
\hskip -0.5 cm
{\Large
$h(\sigma)$}

\newpage
\thispagestyle{empty}
\hspace{5cm}
{\Large
\textbf{
Fig. 5}
, Bellier-Castella et al., JCP}
\\[2cm]
\Large
\begin{center}
\unitlength1mm
\begin{picture}(140,140)
\put(0,0){\makebox(140,140)[b]  
{\epsfysize140mm \leavevmode \epsffile{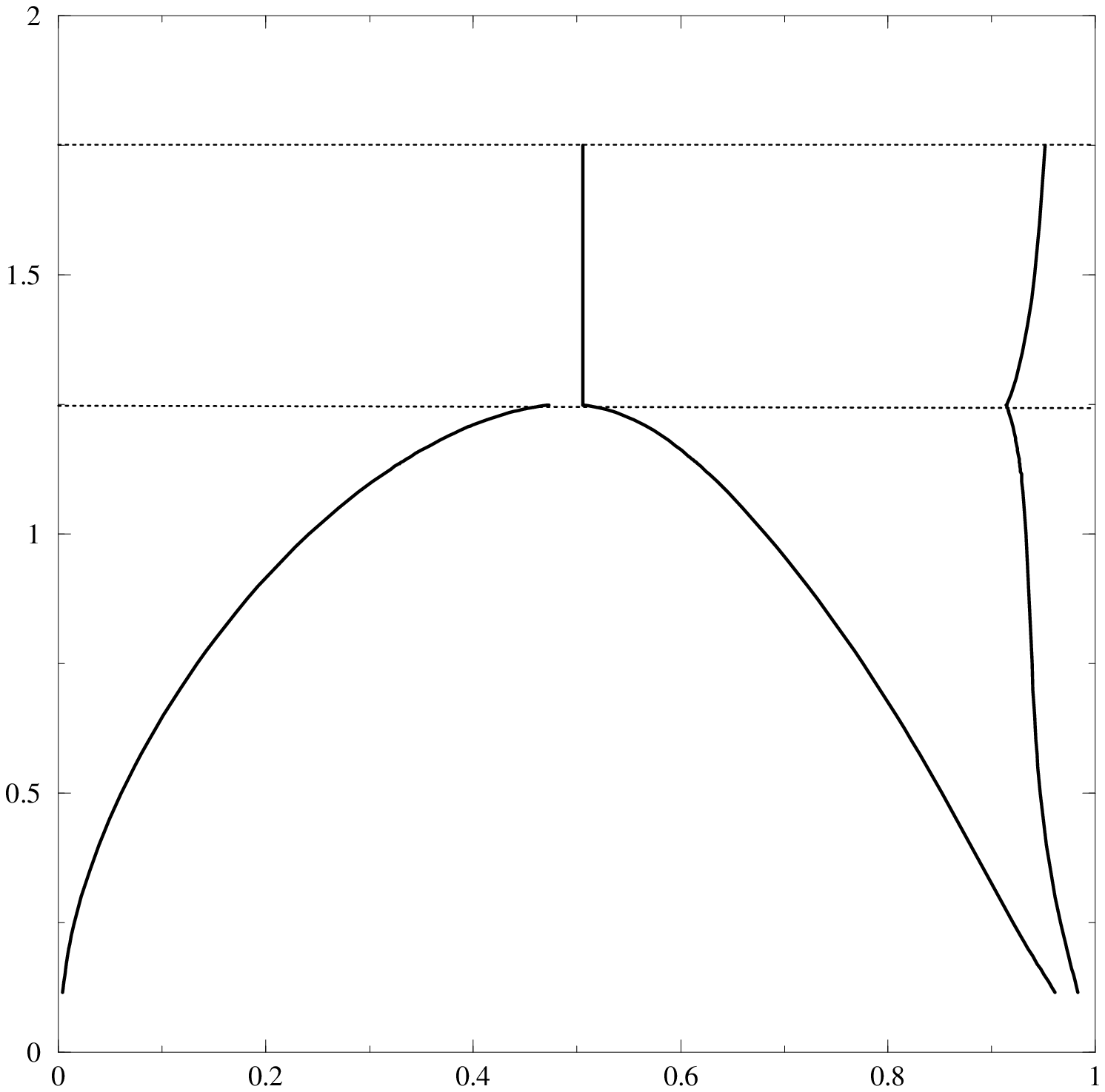} }}
\end{picture}
\end{center}

\vskip -1 cm
\hskip 12 cm
{\Large
$\eta$}

\vskip -12 cm
\hskip 0. cm
{\Large
$t$}
 
\vskip 0. cm
\hskip 0 cm
{\large
$t_{-}$}

\vskip -4.5 cm
\hskip 0 cm
{\large
$t_{+}$}

\newpage
\thispagestyle{empty}
\hspace{5cm}
{\Large
\textbf{
Fig. 6}
, Bellier-Castella et al., JCP}
\\[2cm]
\Large
\begin{center}
\unitlength1mm
\begin{picture}(140,140)
\put(0,0){\makebox(140,140)[b]
{\epsfysize140mm \leavevmode \epsffile{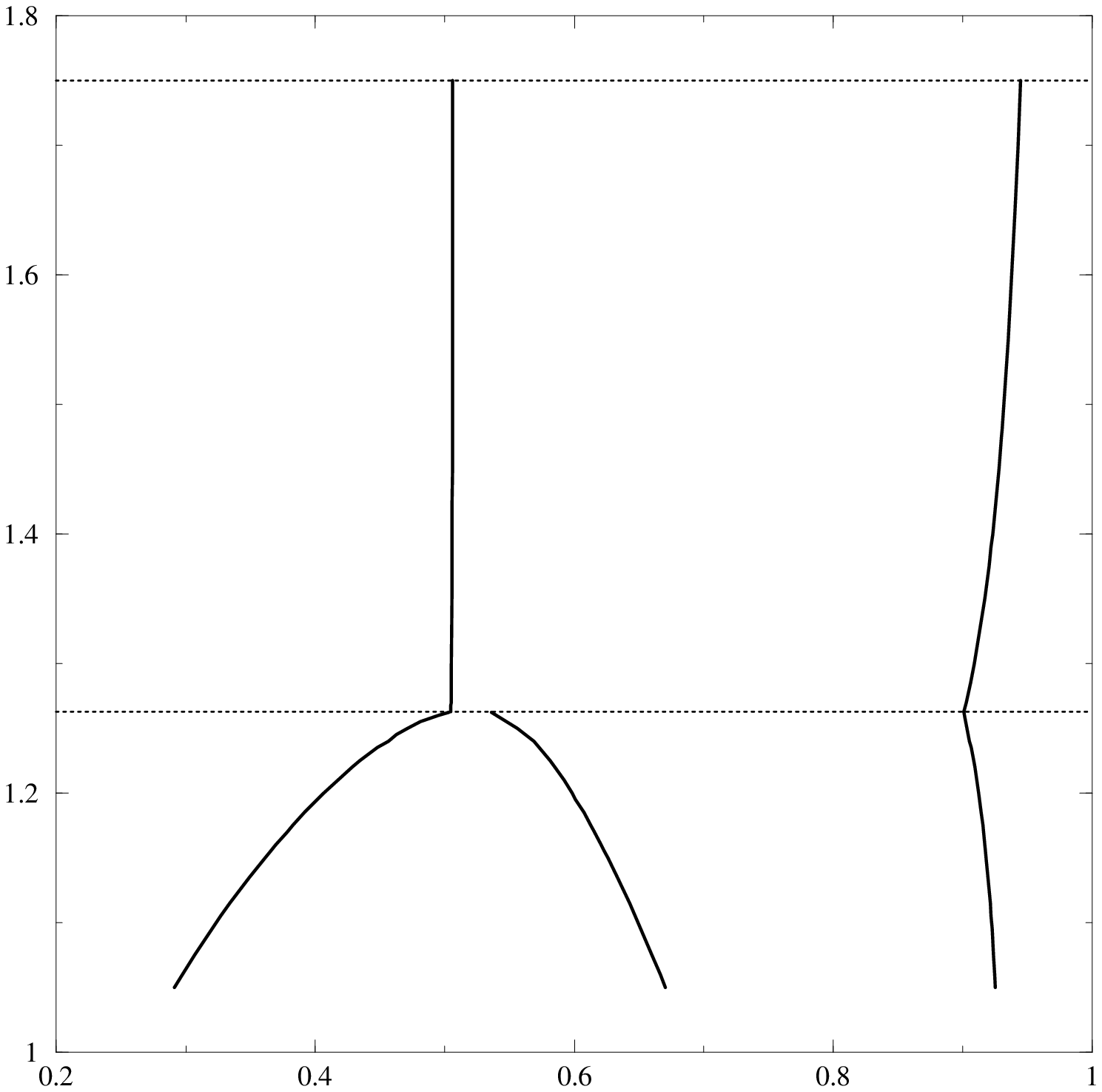} }}
\end{picture}
\end{center}

\vskip -1 cm
\hskip 12 cm
{\Large
$\eta$}

\vskip -12 cm
\hskip 0. cm
{\Large
$t$}

\vskip 4. cm
\hskip 0. cm
{\large
$t_{-}$}

\vskip -9.3 cm
\hskip 0. cm
{\large
$t_{+}$}

\end{document}